\documentclass[%
 reprint,
 amsmath,amssymb,
 aps,
floatfix,
]{revtex4-2}

\usepackage{xcolor}
\usepackage{comment}
\usepackage{graphicx}
\usepackage{dcolumn}
\usepackage{bm}

\begin{document}

\preprint{APS/123-QED}

\title{State-and-rate friction in contact-line dynamics 
}

\author{Chloe W. Lindeman}
 \email{cwlindeman@uchicago.edu}
\author{Sidney R. Nagel}%
\affiliation{
 Department of Physics and The James Franck and Enrico Fermi Institutes \\ 
 University  of  Chicago, \\ 
 Chicago,  IL  60637,  USA.
}

\date{\today}

\begin{abstract}
In order to probe the dynamics of contact-line motion, we study the macroscopic properties of sessile drops deposited on and then aspirated from carefully prepared horizontal surfaces. By measuring the contact angle and drop width simultaneously during droplet removal, we determine the changes in the shape of the drop as it depins and recedes. Our data indicate that there is a force which opposes the motion of the contact line that depends both on the amount of time that the drop has been in contact with the surface and on the withdrawal rate. For water on silanized glass, we capture the experimentally observed behavior with an overdamped dynamical model of contact-line motion in which the phenomenological drag coefficient and the assumed equilibrium contact angle are the only inputs. For other liquid/substrate pairs, the observed contact-line motion suggests that a maximum static friction force is important in addition to damping. The dependence on time of contact and withdrawal rate, reminiscent of rate-and-state friction between solid surfaces, is qualitatively consistent across three substrate-liquid pairs.  
\end{abstract}

\maketitle


\section*{Introduction}

Liquid-substrate interactions can be highly complex. Some of the complexities come into focus for a small liquid drop residing on a solid surface as the drop is removed either by evaporation or by aspiration. The motion of such a drop is mediated by the dynamics of the contact line between the three phases: the liquid, the solid substrate on which it sits, and the surrounding gas. Depending on the circumstances, the drop may respond to the decreasing volume in a variety of ways. In some cases, the contact line slides along the surface as the drop volume decreases so that the angle that the drop surface makes with the substrate, the contact angle, remains constant. In other cases, however, the contact line gets stuck and the drop dries completely without the contact line ever moving. For that to happen, the contact angle has to decrease as the volume of the liquid drops to zero.  In many cases, the contact-line motion is a mixture of these two limiting behaviors.

Contact-line motion is affected by various timescales as well as substrate and fluid properties. Complications occur because the surfaces and liquids are far from ideal, with microscopic details that strongly affect the drop behavior. In this paper, we study the contact-line motion of drops as they are removed from a solid surface by aspiration. We find that the dynamics depend strongly on the time the drop has been in contact with the solid surface (wait time) and on how quickly the drop is removed (flow rate); long wait times or high flow rates give rise to lower contact angles as the contact line recedes towards the drop center. An important conclusion is that the effects of wait time are not confined to the drop's initial contact line but instead appear to be due to changes throughout the entire drop footprint where the liquid makes contact with the solid.  This suggests that there is a dissipative force that increases with the time of contact between the liquid and substrate, reminiscent of friction between solid surfaces. 

To make this quantitative, we construct a model of contact-line motion using overdamped dynamics so that the velocity of the contact line is proportional to the net force it experiences. The inputs to the model are (i) the drag coefficient for motion of the contact line along the substrate and (ii) the equilibrium contact angle, or the angle at which there is no net force pulling or pushing on the contact line. The resulting dynamics, obtained by numerically integrating this model, capture the major features seen in experiments.

We note that there is considerable variation in day-to-day values of essential measurements such as the resting contact angle despite great care in preparing the surfaces in a similar and reproducible manner, and using the same fluids with the same drop volume. This suggests, just as in studies of solid-solid friction, that the damping forces are subtle and controlled by microscopic chemical/physical properties of the interface between the liquid and substrate. Such variations, we believe, have been under-reported in the literature on contact-line depinning and contact angle measurements.  These fluctuations can be large relative to the changes we measure. Nevertheless, we show that the trends described are consistent not only across samples but across different solid-liquid pairs, suggesting that the effects of wait time on liquid-solid friction are robust.

\section*{Background}

The behavior of sessile drops has been studied since 1805, when Young first argued that drops maintain a static ``equilibrium'' contact angle with the surface on which they sit~\cite{young1805}. In that formulation, the contact angle is determined by horizontal balance between the forces created by three surface tensions between the liquid, solid and gas interfaces.  The forces created by the substrate/liquid and substrate/gas surface tensions ($\gamma_{SL}$ and $\gamma_{SG}$ respectively) are co-linear and point tangential to the substrate surface.  At the contact line, the force due to liquid/gas surface tension ($\gamma_{LG}$) points along the surface of the liquid, which forms an angle $\theta$ with the substrate. For force balance to occur in the horizontal direction, the liquid has to assume an equilibrium contact angle, $\theta_{eq}$, such that $$\gamma_{SG} = \gamma_{SL} + \gamma_{LG}~  \cos{\theta_{eq}}.$$ 
The vertical component of the surface tension forces can deform the solid substrate~\cite{jerison2011deformation}.

In practice, drops frequently display a variety of static contact angles. Such variation was originally explained by distinguishing between the macroscopic (measured) contact angle and the microscopic (local) contact angle~\cite{joanny1984}. If there is a defect, for example due to surface roughness, Young's angle may be satisfied locally even as the macroscopic contact angle changes. A similar argument can be made for surfaces with chemical inhomogeneities. However, this does not explain the presence of significant contact angle hysteresis on extremely smooth and clean surfaces. One outstanding example is the case of a superfluid helium drop, which has no impurities, on a freshly deposited surface of cesium.  In that case it was found that the contact line pins completely as the drop is removed~\cite{ross1997}.


A number of approaches, which we will return to in the Discussion section, have been advanced to explain such large contact-angle hysteresis.  It is possible to have a drop with contact angle $\theta \neq \theta_{eq}$ if  the contact line is out of equilibrium; theories about the motion of such contact lines must account for the divergence of shear stress at a moving contact line~\cite{snoeijer2013moving,voinov1976hydrodynamics,blake1969kinetics}.
Contact-line dynamics may also depend on the history of deposition on rough substrates~\cite{quere2008wetting}.

One general approach has been to consider contact-line hysteresis as being due to a form of friction.  Ideas about solid-solid friction --- a microscopic effect that is extremely complex yet is described in some cases by models simple enough to be taught in introductory physics classes --- are promising, and friction has appeared in the description of liquids moving across solid surfaces for over half a century~\cite{yarnold1938motion,bikerman1950sliding,furmidge1962,carlson2012,gao2018,barrio2020contact}.

Much of the past work measuring the adhesion force of the contact line has involved drops whose radial symmetry is broken by gravity, capillary forces, or effective forces due to rotation~\cite{tadmor2008, gao2018, tadmor2009}. In our studies, the drop remains approximately axisymmetric throughout the experiment. Because of the simplified geometry, only the receding (dynamic) contact angle is relevant. Most importantly, the drop footprint in these experiments is always contained within the area defined by the initial contact line, distinct from drop-sliding experiments where the behavior may depend on whether or not the drop has left its initial area of contact~\cite{furmidge1962}. 

\section*{Results}

\subsection*{Water on silanized glass}

\begin{figure}
\centering
\includegraphics[width=8.6cm]{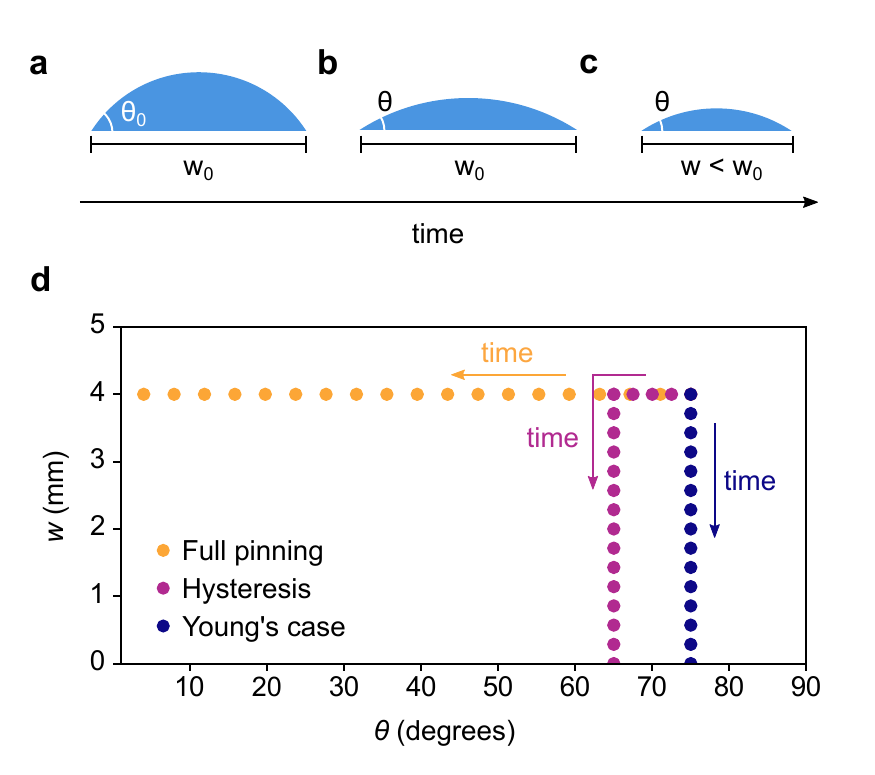}
\caption{
Schematic of experimental protocol and analysis.  (a) A drop has just been put down; aspiration begins at a fixed flow rate. (b) Initially, the drop is pinned: the contact angle decreases while the drop width stays fixed at $w=w_0$. (c) Eventually, the drop depins and the drop width begins to decrease so $w<w_0$. (d) Schematic drop width versus contact angle for three idealized situations. The blue points correspond to the case where there is no pinning; the contact angle remains fixed at the equilibrium angle so that the data  would appear as a vertical line of points (constant contact angle). The yellow points show the limiting case of full pinning: the data would appear as a horizontal line of points (constant width). The purple points show what is expected in the case of partial pinning, where the drop originally remains at a fixed width until it depins and then recedes with constant contact angle. Time during withdrawal is shown by arrows, moving to the left and down as the drop volume decreases.
}
\label{cartoon}
\end{figure}

We measure the contact-line motion of ultrapure water drops on a flat horizontal silanized glass surface as the liquid inside the drops is removed. Surfaces were carefully cleaned and prepared as discussed in the Methods section below. Water on silanized glass has only moderate contact-angle hysteresis so that the contact line is initially pinned during aspiration but depins well before the drop is fully removed. We allow the drop to sit for a prescribed wait time $\tau$ in a nitrogen environment before removing it via aspiration through a thin tube at a constant flow rate $q$. During the withdrawal process, we measure two macroscopic parameters: the contact angle, $\theta$, and the width, $w$, of the drop at its base, as depicted in Fig.~\ref{cartoon}. 

\begin{figure}
\centering
\includegraphics[width=8.6cm]{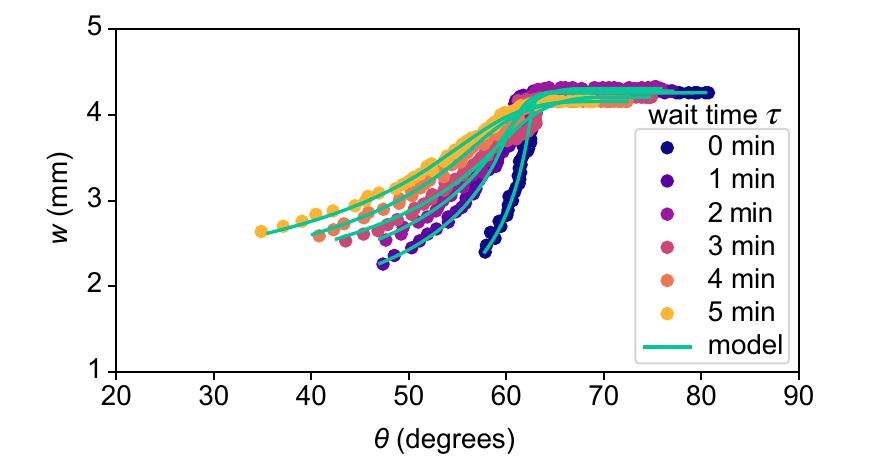}
\caption{
Drop width, $w$, versus contact angle, $\theta$, during aspiration of water drops on silanized glass for the different wait times, $\tau$, shown in the legend. After being in contact with the surface for time $\tau$, each drop is then withdrawn at fixed flow rate, $q = 2$ $\mathrm{\mu}$L/s. The best-fit curves from the model are shown as the solid lines. 
}
\label{wait_time}
\end{figure}

In one type of experiment, shown in Fig.~\ref{wait_time}, we vary the wait time, $\tau$, between $0$ and $5$ min while keeping the flow rate fixed at $q = 2$ $\mathrm{\mu}$L/s. Drops which have sat for longer times tend to have lower dynamic contact angles at a given width than drops put down and immediately removed. In other words, for $\tau = 0$ the changing volume of the drop is accommodated primarily by a change in its width leaving the contact angle relatively constant; at larger $\tau$, on the other hand, the volume change creates a more pronounced change in the contact angle (and therefore a less pronounced change in the width).



\begin{figure}
\centering
\includegraphics[width=8.6cm]{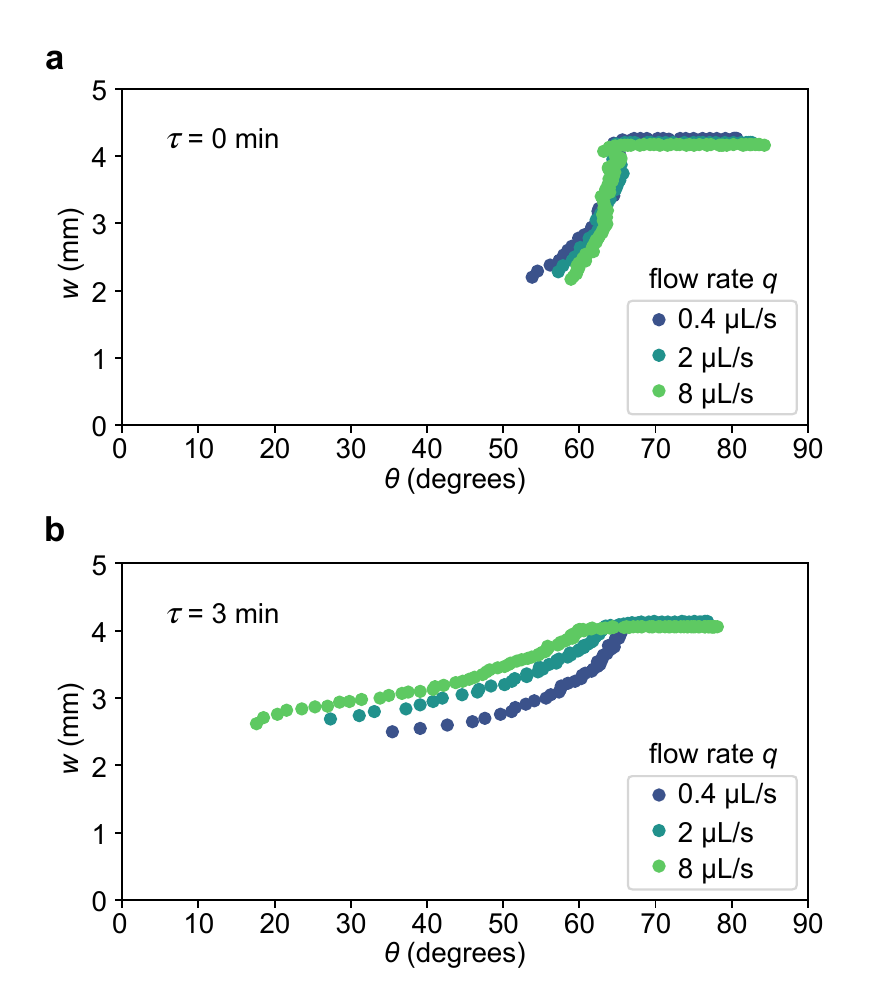}
\caption{
Drop width, $w$, versus contact angle, $\theta$, during aspiration of water drops on silanized glass for the different withdrawal flow rates, $q$, shown in the legend.  (a) When the drops are removed immediately after deposition, $\tau \approx 0$, the curves at all $q$ look similar. (b) When drops sit for $\tau = 3$ minutes before withdrawal, higher $q$ leads to lower depinning and dynamic contact angles.
}
\label{flow_rate}
\end{figure}

In another type of experiment, shown in Fig.~\ref{flow_rate}, we vary $q$ between $0.4$ and $8$ $\mathrm{\mu}$L/s with $\tau$ kept fixed at either $\tau = 0$ or $3$ minutes.  The effect of flow rate is more subtle than the effect caused by a change in waiting time. Drops deposited and then removed immediately show little dependence on withdrawal rate. However, for drops with longer wait times ($\tau = 3$ minutes), higher withdrawal rates lead to lower depinning angles. 

We modify the original experiment to remove the effect of pinning at the drop's initial contact line.  In these modified experiments, after roughly half of the drop volume is removed, the drop is allowed to relax at fixed volume for several seconds, which is long enough for the contact line to come to rest. Only after the drop is relaxed in this state is the remainder of the drop aspirated at the same flow rate as in the first part of the aspiration process. In the first step of these modified experiments, enough liquid is removed that the drop depins from its initial contact line, effectively resetting the ``initial conditions'' of the experiment. Thus the initial contact line does not play a role in the drop's subsequent motion after restarting the aspiration. 

\begin{figure}
\centering
\includegraphics[width=8.6cm]{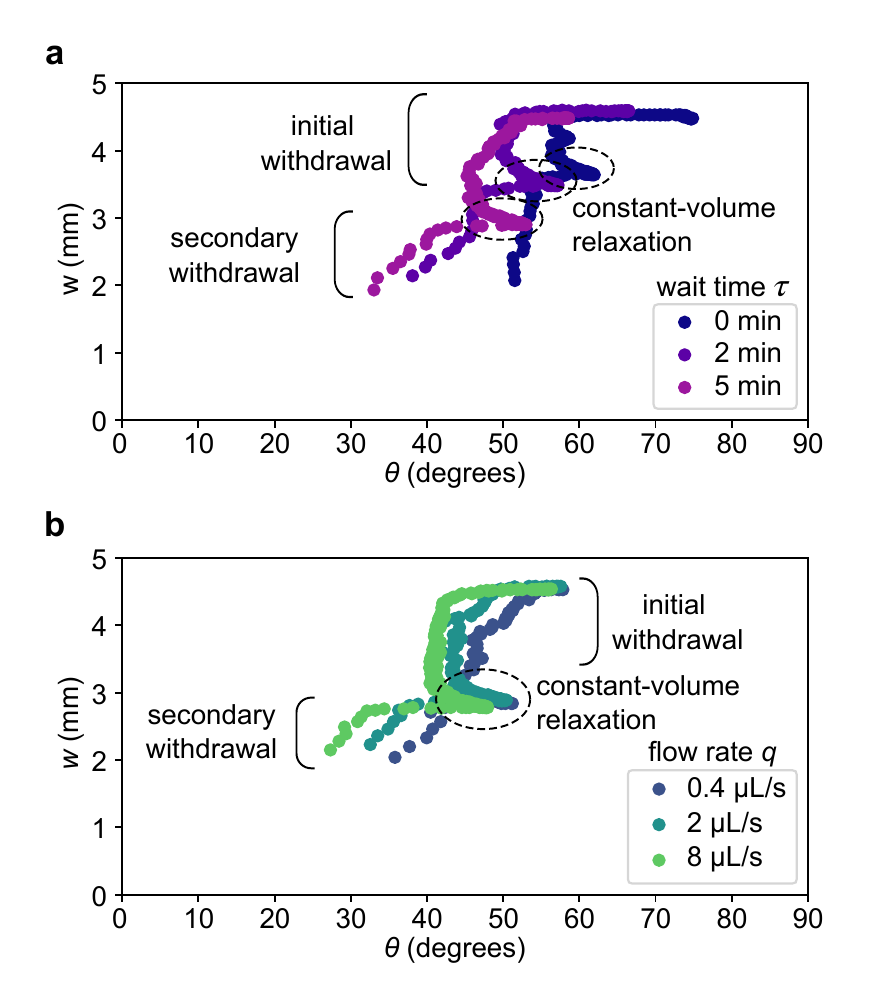}
\caption{
 For water on silanized glass, drop width, $w$, versus contact angle, $\theta$, during paused aspiration of drops. The dashed ovals indicate regions where aspiration was halted, allowing the drop to relax to a static configuration. Experiments done at (a) different wait times, shown in legend (fixed $q=2$  $\mu$L/s) and (b) different $q$ in legend (fixed $\tau=5$ minutes). Higher flow rates and longer wait times correspond to lower contact angles even after the drops have depinned and are allowed to relax.
}
\label{relax}
\end{figure}

The drops relaxed during the constant-volume phase by increasing their contact angle while decreasing their width.  This results in non-monotonic behavior in $w$ vs. $\theta$ as shown in Fig.~\ref{relax}. After restarting the flow, the behavior of the drop quickly returns to its pre-relaxation trajectory, dependent on the initial wait time and flow rate.  This was the case even though the time allowed for the drop to relax was the same in all experiments.

Figure~\ref{relax}a shows that the change in behavior with wait time is dominated by an effect from the entire region where the drop was in contact with the substrate (the drop's initial footprint), not only the edge of that region (the initial contact line). Figure~\ref{relax}b demonstrates that even drops removed at the slowest flow rate relax when the flow is turned off. This indicates that none of the experiments reported here can be considered quasistatic, with the contact line in approximate force balance at all times, since quasistatic behavior would require the contact line to stop moving as soon as the flow is turned off.

\subsection*{Model with overdamped contact-line dynamics}
\label{sec:model}

We model the contact-line behavior as the drop volume decreases linearly in time using four assumptions:
\begin{enumerate}
    \item Drops maintain a spherical cap geometry throughout, so that the volume is a function only of the contact angle and drop width. This corresponds to an assumption that the drop is small enough that the effects of gravity are negligible ($w$ not much larger than $(\gamma_{LG} / (\rho g))^{1/2}$, where $\rho$ is the liquid density and $g$ is the acceleration due to gravity) and that the withdrawal rate is sufficiently slow that it does not distort the drop surface.
    \item The force per unit length on the contact line is given by $\gamma_{LG}(\cos{\theta} - \cos{\theta_{eq}}),$ where $\gamma_{LG}$ is the surface tension of the liquid in air, $\theta$ is the instantaneous value of the contact angle, and $\theta_{eq}$ is the equilibrium contact angle.
    \item The contact line moves according to $F = \beta v$, where $\beta$ is a damping coefficient and $v$ is the velocity of the contact line, equivalent to $\partial_\mathrm{t}(w/2)$.
    \item The drop can only recede, not spread.
\end{enumerate}
Note that to describe water on silanized glass, we did not find it necessary to include any static friction term, as there would be for solids moving on solids. 

For each experiment, the model was integrated numerically using the measured initial contact angle and width as well as the calculated experimental flow rate (see Methods for details). For each experiment, we determined the best fit values of $\theta_{eq}$ and $\beta$. 

\begin{figure}
\centering
\includegraphics[width=8.6cm]{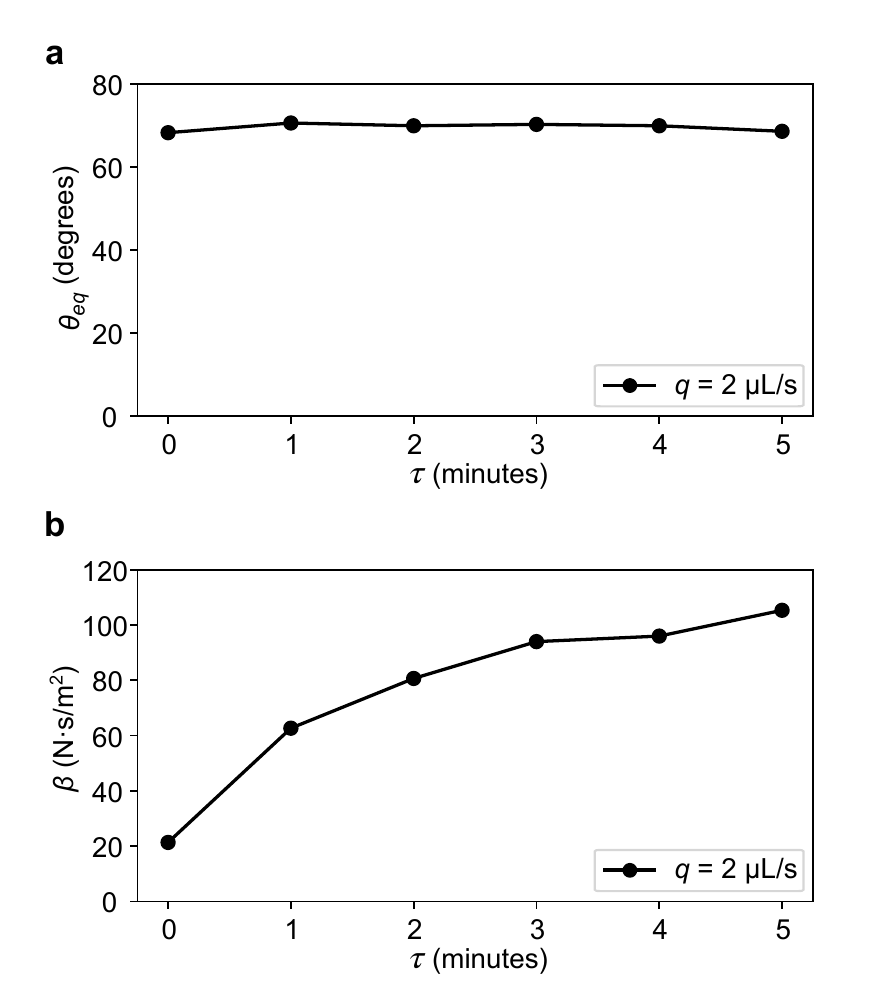}
\caption{
Model fit parameters for water on silanized glass showing (a) equilibrium contact angles, $\theta_{eq}$, and (b) damping coefficients, $\beta$, versus wait time, $\tau$, for fixed flow rate $q=2$ $\mu$L/s. Values shown are averages over two or three sets of data taken under the same experimental conditions.
}
\label{fit_results}
\end{figure}

\begin{figure}
\centering
\includegraphics[width=8.6cm]{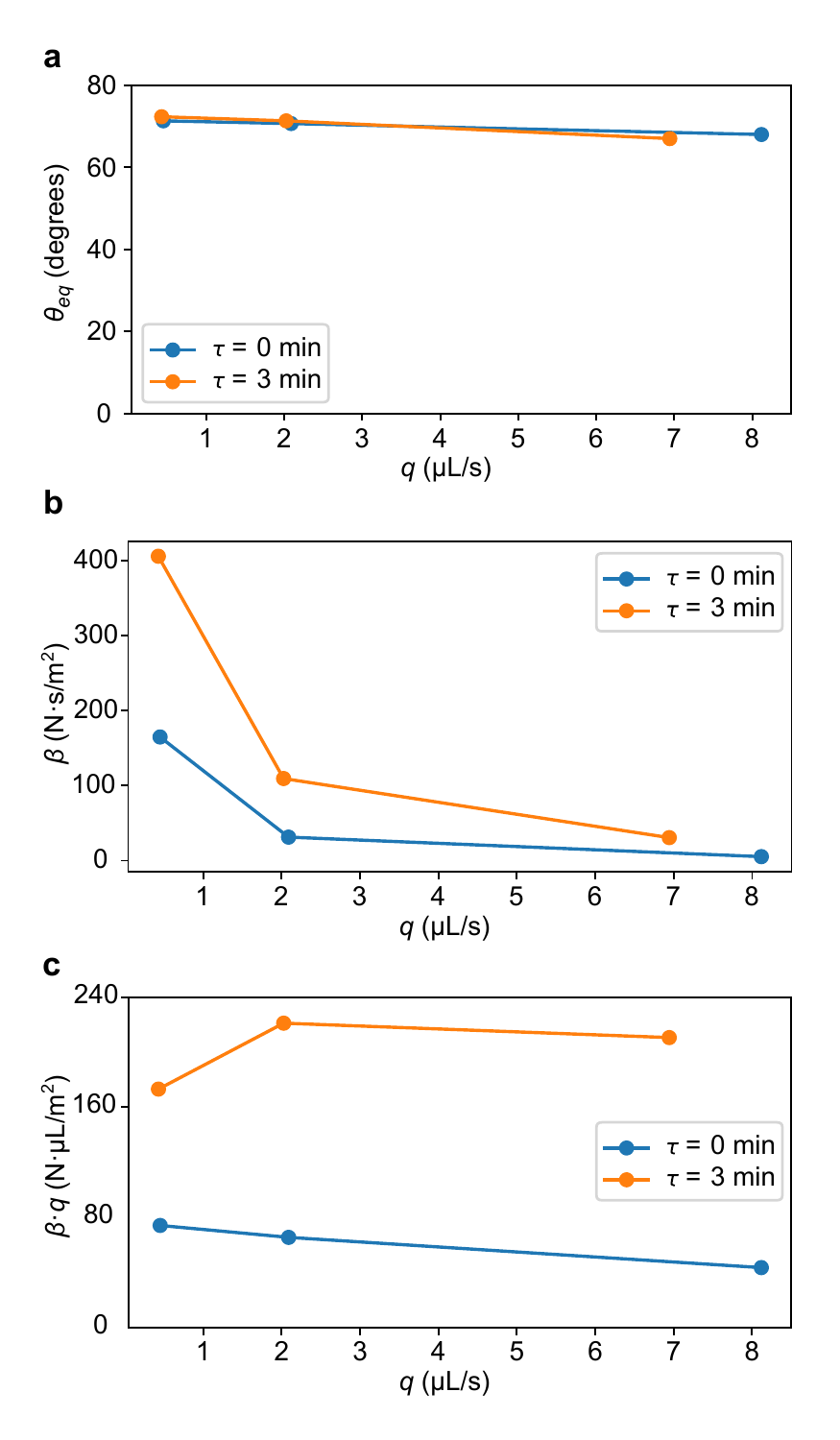}
\caption{
Model fit parameters for water on silanized glass showing (a) equilibrium contact angles, $\theta_{eq}$, and (b) damping coefficients, $\beta$, 
versus flow rate, $q$. (c) The damping coefficients from (b) multiplied by the flow rate are plotted versus $q$. 
Values shown are averages over three sets of data taken under the same experimental conditions. 
}
\label{fit_results_q}
\end{figure}

Figure~\ref{wait_time} shows the best-fit results from the model overlaid on the experimental data. The fits capture the essential features of the experiments, including initial pinning, depinning, and continued decrease in both contact angle and width. The average best-fit values of $\theta_{eq}$ and $\beta$ are plotted in Figs.~\ref{fit_results} and~\ref{fit_results_q} for the experiments varying wait time, $\tau$, and flow rate, $q$, respectively. 

Figures~\ref{fit_results}a and~\ref{fit_results_q}a show that the equilibrium contact angle does not vary appreciably as a function of either $\tau$ or $q$.  The results for the damping coefficient, $\beta$, however show significant variation with both  $\tau$ and $q$.  In particular, Fig.~\ref{fit_results_q}b shows that the behavior of $\beta$ drops dramatically with increasing $q$. In Fig.~\ref{fit_results_q}c we multiply the damping coefficients by the flow rate, plotting $\beta q$ versus $q$. This removes most, but not all, of the variation.  We explore this in further detail in the Discussion section below.

\subsection*{Other solid-liquid pairs}

\begin{figure}
\centering
\includegraphics[width=8.6cm]{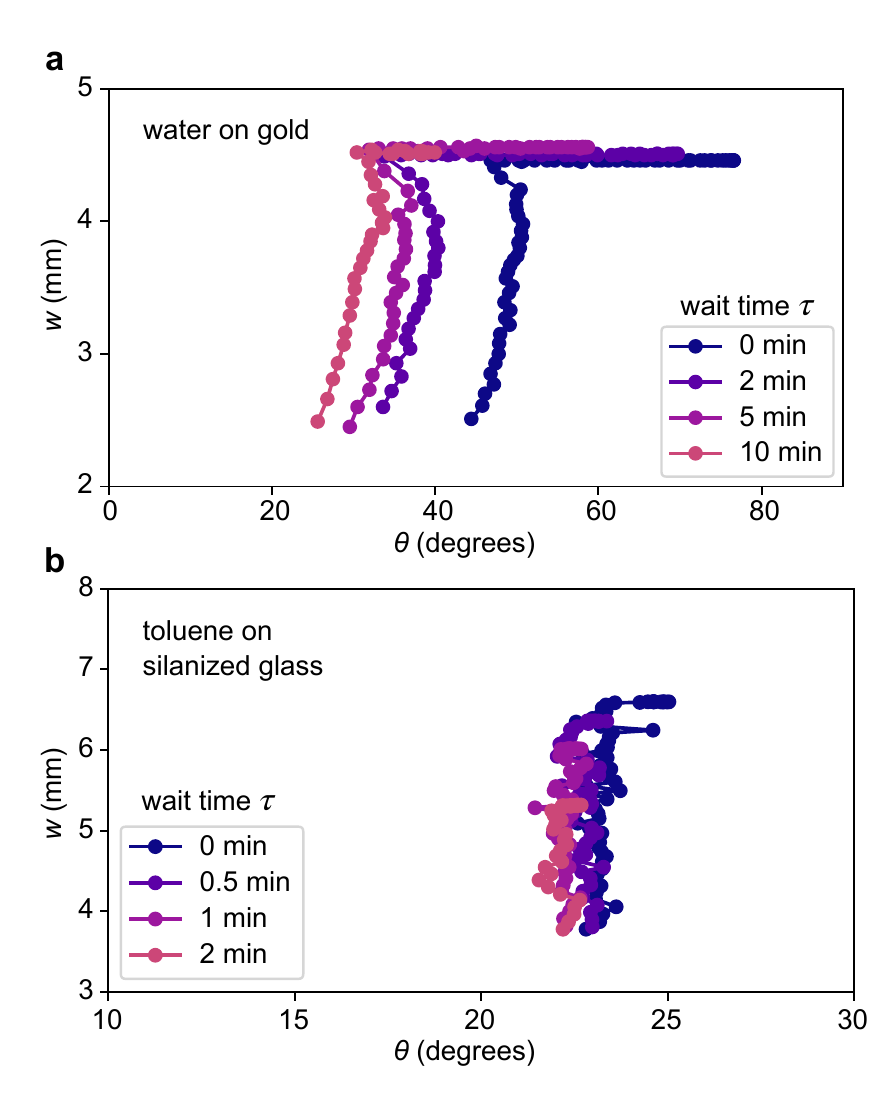}
\caption{
Drop width, $w$, versus contact angle, $\theta$, for (a) water on freshly deposited gold and (b) toluene on silanized glass. Legends show the wait times, $\tau$, for each experiment.   The withdrawal flow rate was fixed at $q = 2$ $\mu$L/s.
}
\label{other_pairs}
\end{figure}

We repeated the wait-time and flow-rate experiments with two other solid-liquid pairs: water on evaporated gold and toluene on silanized glass, shown in Figs.~\ref{other_pairs}a and b respectively. The details of the drop dynamics vary substantially from case to case. Compared with water on silanized glass, water on gold depins more abruptly and remains pinned to lower contact angles. Toluene on silanized glass starts with a much lower contact angle than water on either surface. 

Nevertheless, despite these variations in quantitative and even qualitative behavior, the trends observed as a function of $\tau$ are consistent across all three liquid-solid pairs: longer wait times result in lower contact angles after depinning. See the Supplemental Information for the full set of experimental results on all three solid-liquid pairs.

Some of the features observed, for example the increase in contact angle after depinning seen in Fig.~\ref{other_pairs}a, make the assumptions used in the model outlined above with only two parameters (the damping coefficient and the equilibrium contact angle) poorly suited for these solid-liquid pairs. However, the model gives a better fit to the data with an additional assumption and parameter: a maximum force which must be overcome before the contact line will begin to move. We explore this more fully in the Discussion section below.

\begin{figure}
\centering
\includegraphics[width=8.6cm]{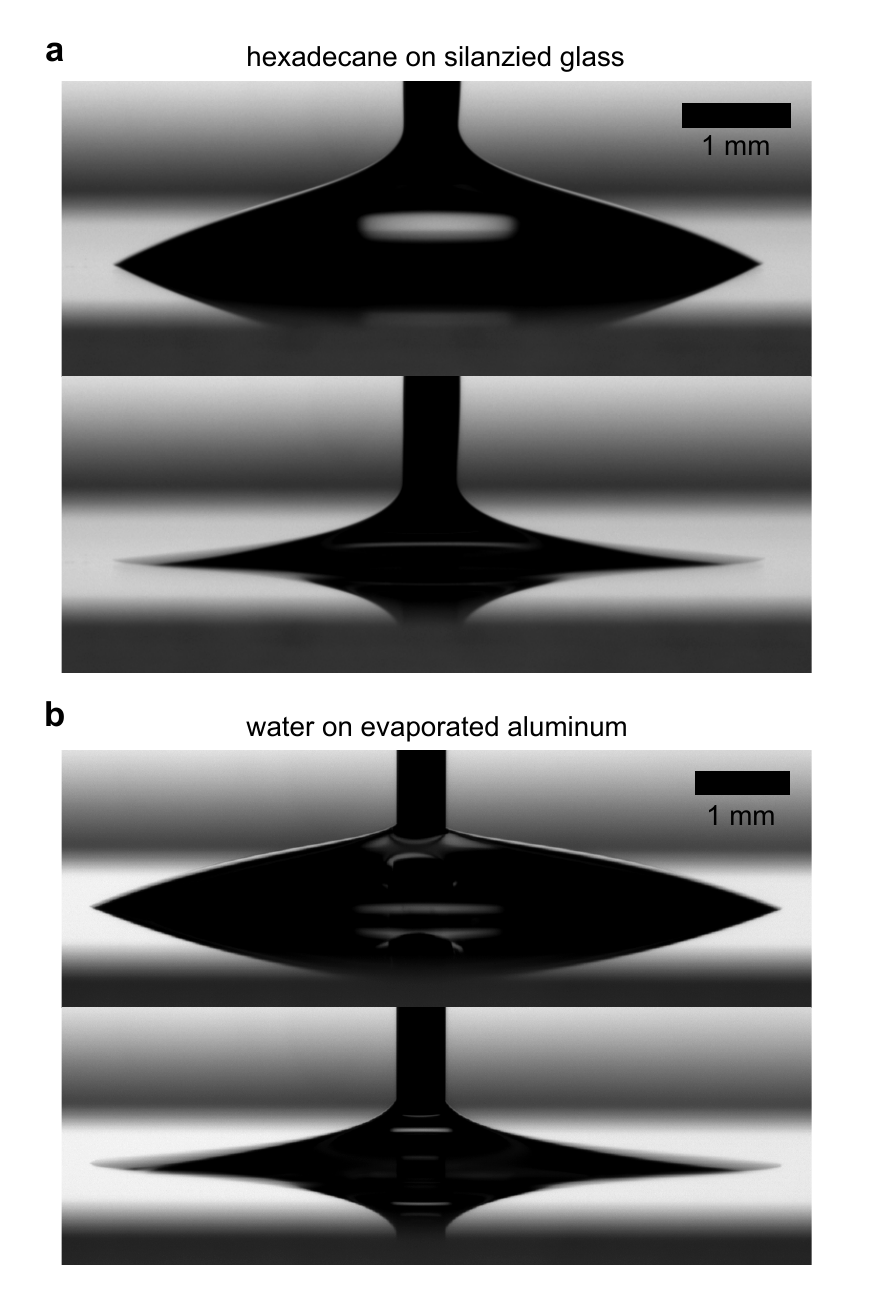}
\caption{
Pure liquids showing complete pinning.
Photographs show the beginning (top) and end (bottom) of aspiration of a drop of (a) hexadecane on silanized glass and (b) water on freshly evaporated aluminum.  In both cases, the contact line has not moved appreciably even after the contact angle has gone to zero.  The withdrawal tube is shown emerging from the top of the drop in each image.
}
\label{pinning}
\end{figure}

Finally, we report two clean systems exhibiting maximum contact-angle hysteresis: pinning down to $\theta=0$. These two systems are water on evaporated aluminum and hexadecane on silanized glass. Representative images from experiments with these two systems are shown in Fig.~\ref{pinning}. Although common in everyday situations where surfaces are rough and drops may contain solute, reports of such extreme contact-angle hysteresis in clean systems with smooth surfaces are rare with a few exceptions~\cite{ross1997,bormashenko2011}.

\subsection*{Reproducibility}
\label{sec:reproducibility}

To study the effect of the wait time and flow rate on contact line dynamics, we would ideally fix all other parameters and vary only $\tau$ and $q$ separately. However, there are a large number of factors that affect the behavior of drops, and in practice not all of them can be precisely controlled. 

We found considerable inconsistency in equilibrium contact-angle values from day to day for all liquid-solid pairs considered. Although the surfaces were carefully cleaned and prepared as consistently as possible (see Methods section), and although the trends reported above remained present, the observed values of the contact angle varied by as much as $15$ degrees. While some substrate-liquid pairs may be more quantitatively reproducible, we suspect that large variations are common and have not been sufficiently reported in the literature.

\begin{figure}
\centering
\includegraphics[width=8.6cm]{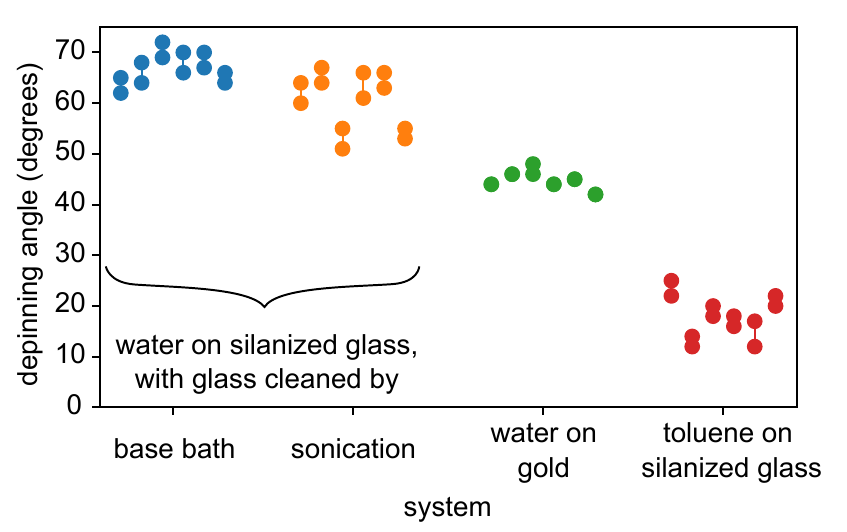}
\caption{
Six different measurements under the same experimental conditions of depinning angles for water on silanized glass prepared with cleaning by base bath, water on silanized glass prepared with cleaning by sonication and plasma etching, water on gold-coated glass prepared with cleaning by sonication and plasma etching, and toluene on silanized glass prepared with cleaning by sonication and plasma etching. Each measurement is shown as upper and lower bound on actual value; data are separated horizontally for visual clarity only. In all cases, the spread for a given system was substantial (between 10 and 15 degrees, much larger than the uncertainty for a single measurement).
}
\label{inconsistent}
\end{figure}

Figure~\ref{inconsistent} shows the experimental scatter of results for a variety of systems with different substrate cleaning methods, substrate coatings, or liquids. The angle intervals shown give an estimate of the angle at which the drop began to recede from its initial contact line (the ``depinning angle'') for fixed $\tau$ and $q$. Note that while water on gold has a fairly consistent depinning angle, the contact line behavior after depinning was more variable for both water on gold and toluene on silanized glass, with not only quantitative but also qualitative changes from day to day. See the SI for full set of experiments.

While the variation in angle is less than the contact angles themselves (\textit{e.g.}, $10^\circ$ versus $70^\circ$), in all cases it is a significant variation.  More important is that the changes we observe as a function of wait time and flow rate are smaller and would be difficult to observe if we compared results obtained on different days. The relative consistency of measurements on a single day made it possible to observe these trends directly, and we see from the model best-fit plots that averaging does not wash out the observed structure. That is, even through there is a large day-to-day variation of the depinning angle, the trends as a function of $\tau$ and $q$ persist and are fairly consistent for all three solid-liquid pairs.

Neither thermally annealing silanized slides before use nor removing drops from below via a hole in the substrate made a substantial difference in the consistency of the measurements. Likewise, small variations in the needle height during deposition and aspiration did not change the observed contact angles by enough to account for the day-to-day variation, nor did changing the order in which the experiments were done after slide preparation. The inconsistencies must therefore come from sensitivity to a more subtle aspect of the slide preparation or liquid deposition process, for example the precise vacuum pressure during deposition or the humidity of ambient air while transporting slides from one apparatus to another. They therefore seem to be a fundamental feature of these three-phase contact-line measurements.

\section*{Discussion}

We have focused here on nominally pure fluids without solute. While a great deal of attention has been paid to the dynamics controlled directly at the contact line, our results point to the importance of considering the change in properties of the entire surface where the drop is in direct contact with the substrate. The experiments reported here show that both wait time and flow rate --- in other words, both the timescale of contact between the liquid and solid and the timescale of liquid withdrawal --- impact the dynamics of a drop as it is removed. 

\begin{figure}
\centering
\includegraphics[width=8.6cm]{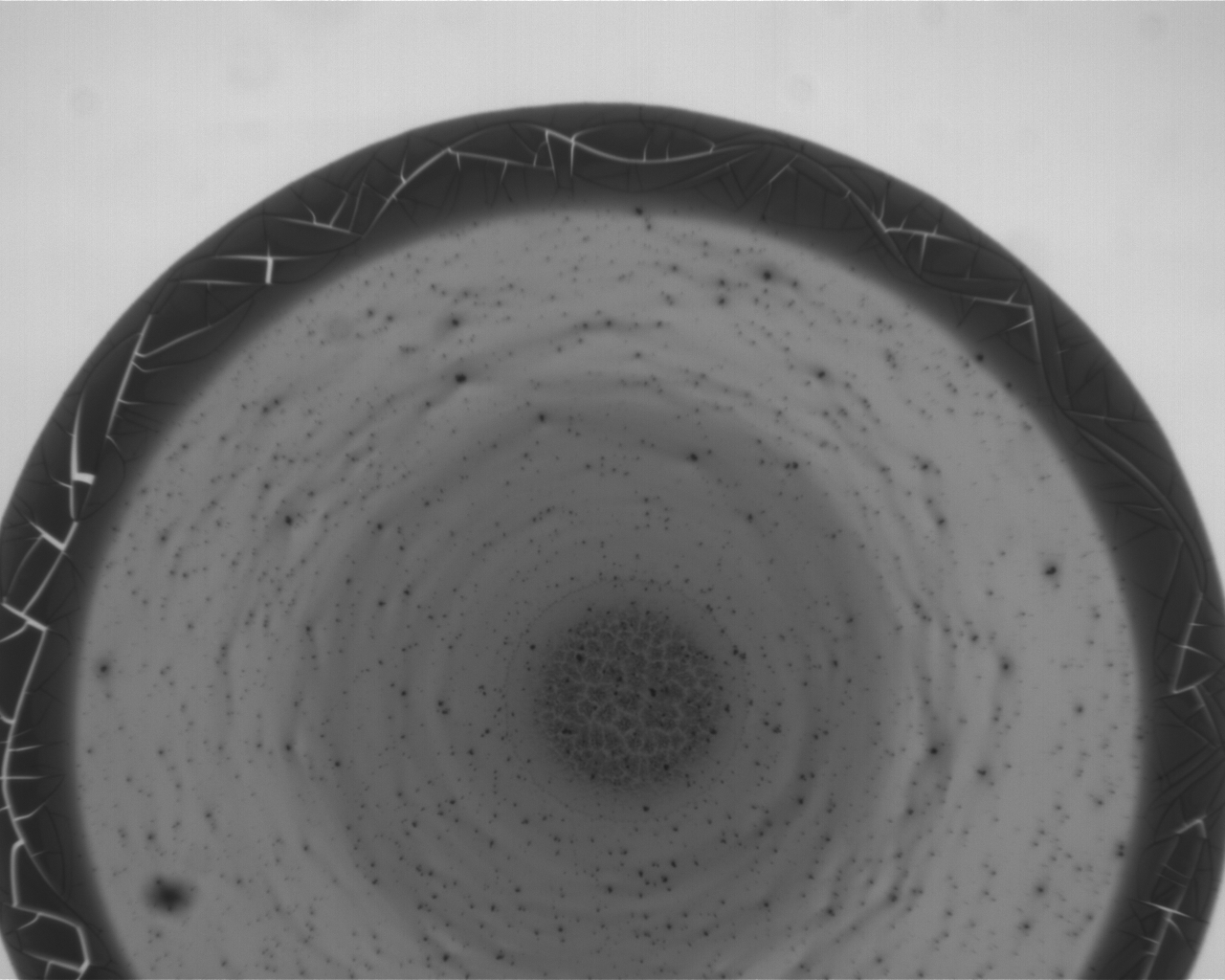}
\caption{
The stain left behind by a 5 $\mu$L drop of Brilliant Blue G in water (mass fraction of $2.5 \times 10^{-3}$) deposited on acrylic. Along with a dark ring stain at the edge of the stain, a dark region is visible in the center of the stain.
}
\label{middlestain}
\end{figure}

As demonstrated in Fig.~\ref{relax}, changes to the dynamics are a result of changes throughout the entire area of contact between the drop and the solid surface, not as has been previously suggested~\cite{tadmor2008} a result of changes at the drop's resting contact line. In a similar vein, the so-called coffee-stain effect has drawn attention to the fact that, due to evaporation from the edges of the drop, any solute is predominantly deposited just inside the contact line~\cite{deegan1997,deegan2000contact}. Additional studies have shown that this effect can be suppressed by changing the geometry of the particles in solution~\cite{yunker2011suppression} or the surface activity of the solute~\cite{li2013evaporation}. We note here that even in simple cases with microscopic solute and a well-defined contact-line deposit, there is often a large deposit of solute that remains near the center of the drop. This has often been neglected. An example is shown in Fig.~\ref{middlestain} for Brilliant Blue G in water. Both of these observations indicate that the interior of the drop, where the liquid makes contact with the surface, contributes in important ways to the overall dynamics of pinning and solute deposition. 

This distributed effect is also consistent with the result that a dissipative force acts to oppose motion of the contact line even once it has started to move. 
The analogy to solid-on-solid friction is especially striking when considering the work of Dietrich on rate-and-state friction~\cite{dieterich1979}, which showed that the frictional force between two solid surfaces can depend on both how quickly the surfaces are moving with respect to one another and how long the surfaces have been in contact. Here we have also observed also a state effect: the increase in the damping coefficient as wait time, $\tau$, increases. The damping coefficient, $\beta$ depends strongly on $q$ so that $\beta q \approx$ constant.  

Moreover, the behavior seen for other solid-liquid pairs as shown in Fig.~\ref{other_pairs} is reminiscent of other types of solid-solid friction. In cases with a sharp depinning event, where the contact line jumps suddenly to smaller width and larger contact angle, the addition of a maximum static friction term may be necessary to capture the dynamics. Likewise, the obvious non-monotonic behavior seen in some water-on-gold and toluene-on-glass measurements, even after depinning, are suggestive of stick-slip behavior where the contact line overshoots its equilibrium position. In this case, the acceleration term (implicitly dropped in the overdamped description of motion) can no longer be neglected.

We have so far been agnostic as to the cause of dissipation leading to the change of behavior with wait time. However, we can rule out certain effects and suggest others as possibilities. As proposed by Young, the basis for our analysis is that any net force on the contact line comes from the surface tension of the drop~\cite{young1805}. If, during the course of the drop receding, the contact angle changes substantially, then the frictional force must be on the order of the force due to surface tension. 
    
If the contact line is out of equilibrium, the contact angle need not equal the equilibrium value: $\theta \neq \theta_{eq}$. Theories for contact-line motion must account for the divergence of the shear stress at a moving contact line, typically by including a slip length or other microscopic mechanism~\cite{snoeijer2013moving}. One example is the Cox-Voinov law, which balances surface tension with dissipative forces to predict the dynamic contact angle as a function of capillary number~\cite{voinov1976hydrodynamics}. Another is the so-called molecular-kinetic theory, which describes contact-line motion when it is dominated by adsorption and desorption events~\cite{blake1969kinetics}. 

To assess the importance of these effects, we note that for the observed contact-line velocities ($V \approx$ a few mm/s in the fastest cases), the capillary number, $Ca = V \eta/\gamma_{LG} < 10^{-4}$ (where $\eta$ and $\gamma_{LG}$ are the liquid viscosity and liquid-air surface tension respectively). This indicates that viscous forces are small compared to those due to surface tension. In particular, even with the shear stress formally diverging at the contact line, it has been suggested that the total dissipation of a smooth contact line is negligible for $Ca$ less than $\sim 10^{-4}$~\cite{perrin2016}. It thus seems unlikely that the effects of a non-zero slip length are important in our case. We note that inertial forces are also negligible for the observed contact line velocities; for a drop of width $w= 4$ mm, the Weber number $We = \rho V^2 w/\gamma$ (where $\rho$ is the liquid density) is $We < 10^{-3}$.

Within the past decade, the inertial spreading of drops placed on a substrate has been described using analogies to friction for contact-line motion~\cite{carlson2012}. Experiments have shown that drops can have an equivalent of static and kinetic friction as they start to slide over solid surfaces~\cite{gao2018}, and kinetic friction was measured on slippery surfaces by watching a drop's contact angle and width relax from an unstable configuration to a static one after fluid was added or removed~\cite{barrio2020contact}.

Frictional dissipation during contact-line motion on rough surfaces has been attributed to an adhesion force that depends on the surface area wet by the liquid~\cite{quere2008wetting}. When first deposited, microscopic air pockets under the drop may remain in surface declivities so that the actual area of contact is lower than it appears~\cite{cassie1944wettability}. As the drop sits, the contact area increases as the liquid more fully wets the substrate and the drop/substrate adhesion increases ~\cite{wenzel1936resistance}.
In the case of a smooth surface, we propose that the presence of atmospheric gas molecules adsorbed to the surface could play a similar role; the gas dissolves into the water over time. In this case, however, the timescale is set by dissolution and diffusion of gas molecules rather than the Cassie-Wenzel wetting transition.  A mechanism such as this would explain the increase in damping coefficient, $\beta$, that we observe with wait time, $\tau$. 
Interestingly, this mechanism is similar to the idea of deforming contact points that give rise to a changing solid-solid friction in the rate-and-state model~\cite{dieterich1979}.

As suggested above, the effect of flow rate $q$ is more subtle than the effect of wait time. When we vary $q$ while keeping the wait time fixed, neither the equilibrium contact angle, $\theta_{eq}$ nor the rescaled damping coefficient $\beta q$ changes by much as a function of $q$. This may be surprising as Fig.~\ref{flow_rate}b shows curves that are well separated as a function of flow rate. However, comparing the effect of flow rate to the effect of wait time (for example, by comparing Figs.~\ref{flow_rate}a and b), we see that $\tau$ has a much more striking effect, and this, more than the small variation as a function of $q$, is what the results of the model capture. 



In the quasistatic limit, where the behavior is by definition independent of the flow rate, the width versus contact angle curves should be identical; the dynamics are purely determined by force balance, so the only effect of the flow rate is to scale the total time of the experiment. Fig.~\ref{flow_rate}a shows such a case: the curves for different flow rates are nearly identical, suggesting that the experiments are quasistatic. Yet our relaxation measurements (Fig.~\ref{relax}b in particular) indicate that even the slowest of the experiments are not in the quasistatic regime. There must therefore be some other principle at play that causes these curves to lie on top of each other. 

When not rescaled by flow rate, $q$, the damping coefficient, $\beta$, falls dramatically with increasing $q$. In the quasistatic limit, the parameter of interest should be independent of $q$. But the measured damping coefficient $\beta$ would not be independent of $q$ because it implicitly contains the timescale of forcing --- only by dividing out the total time of the liquid withdrawal or, equivalently, multiplying by $q$, would we observe the desired lack of dependence on the flow rate.

In the large body of work on contact-angle hysteresis, a disjoint collection of different factors have been proposed that influence how contact lines move.  The assortment of models that account for these different factors are each valid in a different limit. Friction between solid surfaces exhibits some of the same complexities seen for moving contact lines, with microscopic details that are specific to each liquid/substrate pair studied. Yet a generic description of friction exists: a force proportional to the applied force and opposing the direction of motion. The difficulty in obtaining reproducible equilibrium contact angles on different days, despite great care in the preparation of the substrates, certainly points to the importance of microscopic effects that affect the contact-line dynamics. Nevertheless, we are able to describe some of the macroscopic behavior using a simple model of friction. Additional experiments are clearly necessary to isolate the nature of the microscopic aspects of liquid-solid friction.

\section*{Methods}

\subsection*{Cleaning}

All surfaces were prepared on FisherBrand Plain Precleaned Microscope Slides. Reagents used for preparing slides were 2-propanol ($\ge$99.5\%, Fisher Scientific), acetone ($\ge$99.5\%, Fisher Scientific), sodium hydroxide pellets (Fisher Scientific), ethanol (200 proof, Decon Laboratories), chlorotrimethylsilane (99+\%, Sigma-Alderich), and ultrapure water taken from an Elga Water Purification system.

Most slides were sonicated in acetone for five minutes, rinsed in 2-propanol, rinsed in ultrapure water, and dried with nitrogen before being plasma etched for five minutes under oxygen (100 W, flow rate of 20 cc/min).

In a few cases (water on silanized glass experiments not including the relaxation measurements), slides were cleaned by soaking overnight in base bath and rinsing with reverse osmosis water and then ultrapure water. The base bath was composed of sodium hydroxide, ethanol, and water.

After cleaning, slides were either silanized or coated with metal.

\subsection*{Silanization}

Cleaned slides were placed in a vacuum desiccator. 40~$\mathrm{\mu}$L of chlorotrimethylsilane was placed on one slide (not used for experiments). The desiccator was pumped down for 60 seconds, left for five minutes, then pumped down for another 60 seconds. Four hours later, the slides were removed and cleaned by one minute sonication in acetone, rinsed with 2-propanol, rinsed with ultrapure water, and dried with nitrogen. 

\subsection*{Metal Evaporation}

Cleaned slides were 
loaded in an Angstrom Engineering EvoVac Evaporator. The substrates were rotated at 5 revolutions per minute while deposition occurred. For gold-coated slides, a 7 nm layer of chromium was first deposited, then 50 nm of gold. For aluminum surfaces, 100 nm of alumnimum was deposited. All depositions were done at a rate of 1 $\mathrm{\AA}$/s. 

\subsection*{Experiments}

All slides were used within five hours of preparation. Toluene (99.9\%, Sigma-Alderich) and ultrapure water taken from an Elga Water Purification system were used as the liquids for experiments. All experiments were done in a nitrogen environment (flow rate 15 SCFH) around $25^\circ$C using a Krüss Drop Shape Analyzer (DSA100) with a camera tilt of between 2 and 3 degrees. For each drop, a needle $\sim0.5$ mm in width was positioned $\sim4$ mm above the surface, a 14 $\mathrm{\mu}$L drop was dosed from a syringe connected to the needle, and the needle was moved to a position $\sim0.5$ mm above the surface (inside the drop). The drop was left to sit for wait time $\tau$ before being removed with flow rate $q$.

For relaxations experiments, 10 $\mathrm{\mu}$L were removed (nominal value; actual value was lower because of hysteresis in the removal mechanism). The drop was left to relax for 8 seconds, then the remainder of the drop was removed as before.

Water experiments were analyzed frame-by-frame using the Krüss software Advance with fitting method `Tangent'. Toluene experiments were analyzed using sessile drop analysis software from Github (\url{https://github.com/mvgorcum/Sessile.drop.analysis}) with linear interpolation used to find the drop's edge and a quadratic fit using 14 pixels from the drop's edge; data are reported only down to 2 $\mu$L to minimize the effects of the needle. In all cases, contact angles reported are an average of the left and right contact angles. 

\subsection*{Model}

In the model of the overdamped contact-line dynamics, the volume of a drop was calculated as the volume of a spherical cap with the measured contact angle and width. The initial volume of a drop for a given experiment was set by the measured initial contact angle and width and the volume was decreased linearly over time with a flow rate measured by fitting for the slope of the experimental volume vs time data. For a given equilibrium contact angle, the force per unit length on the contact line at each moment was calculated as $\gamma_{LG} (\cos{\theta} - \cos{\theta_{eq}})$, where $\gamma_{LG}$ is the surface tension of the liquid in air ($0.072$ N/m for water in air). The drop was not allowed to spread (\textit{i.e.}, the width $w$ did not increase even if the force was negative), but if the force was positive the width was decreased according to $$v_{CL} = \frac{|\dot w|}{2} = \frac{F_{net}}{\beta}.$$ 
The final volume of the drop was calculated from the final values of the contact angle and width determined in the experiment, and the numerical iteration of the model was stopped when the drop reached that volume.

For each experiment, the numerical integration of the model was repeated over a wide range of $\theta_{eq}$ and $\beta$ values. The resulting width versus contact-angle curve with the smallest residuals was chosen as the best fit. 

\vspace{.5 cm}
\section*{Acknowledgements}

We thank Tom Witten, Nicholas B. Schade, Zhaoning Liu, Alice Pelosse, Michelle Driscoll, and Thomas Videb$\ae$k for useful discussion. We are also grateful to Margaret Gardel for use of her lab's facilities, Justin Jureller for his instrumentation expertise, and Philippe Bourrianne and Brian Seper for experimental protocols. This work was supported by the US Department of Energy, Office of Science, Basic Energy Sciences, under Grant DE-SC0020972. This work made use of the shared facilities at the University of Chicago Materials Research Science and Engineering Center and the Pritzker Nanofabrication Facility of the Pritzker School of Molecular Engineering at the University of Chicago, which receives support from Soft and Hybrid Nanotechnology Experimental (SHyNE) Resource (NSF ECCS-2025633), a node of the National Science Foundation’s National Nanotechnology Coordinated Infrastructure. C.W.L. was supported in part by a National Science Foundation Graduate Research Fellowship under Grant DGE-1746045.

\section*{Supplemental Information}

\begin{figure*}
\centering
\includegraphics[width=17.2cm]{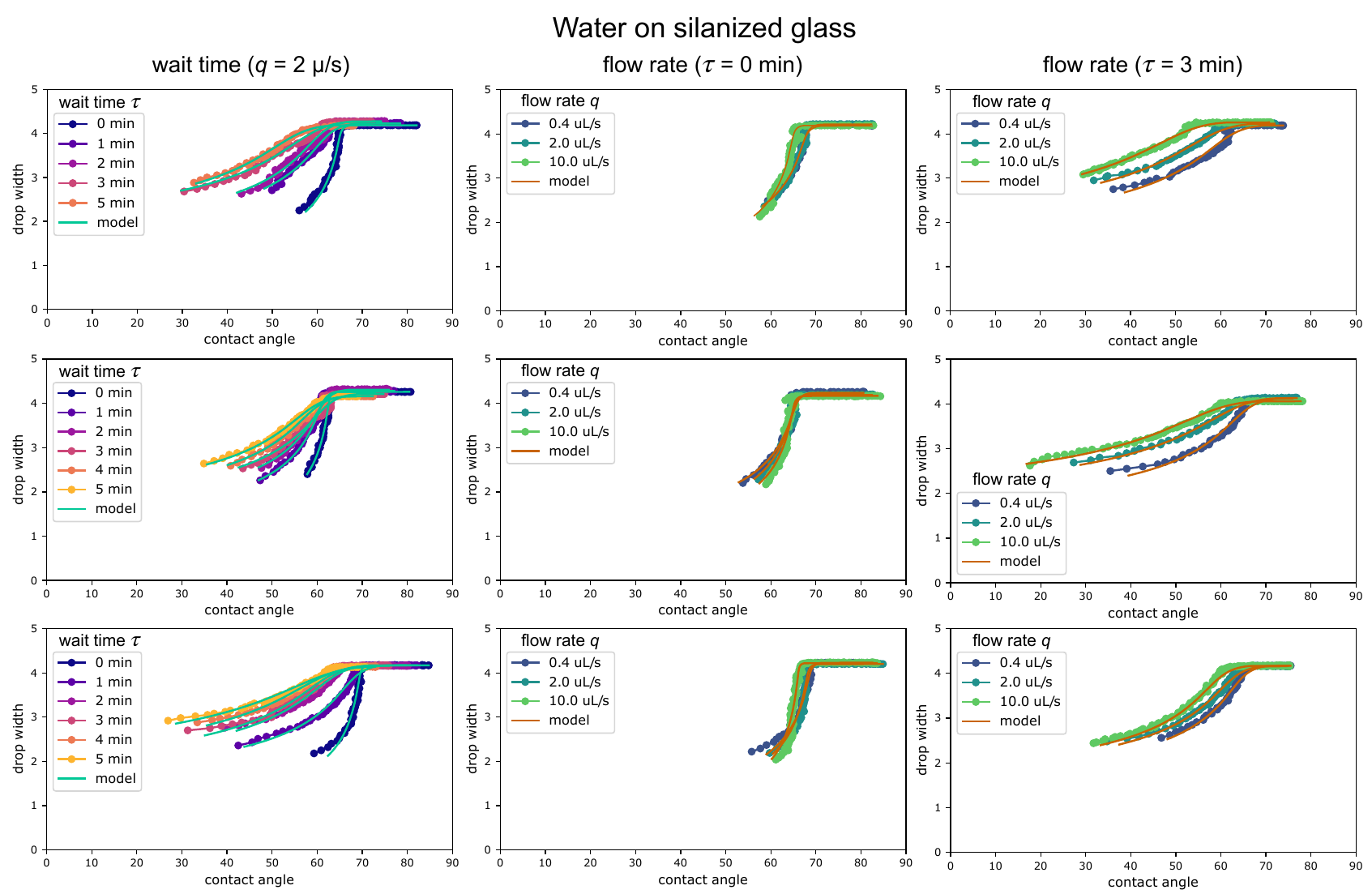}
\caption{
Drop width, $w$ versus contact angle, $\theta$ for water on silanized glass.  The best-fit model curves are overlaid.  Left column shows data at the wait times, $\tau$, in the legend with $q = 2$ $\mu$L/s.  The middle and right columns show data at the flow rates, $q$, given in the legend.  Middle column shows data for $\tau \approx 0$ minutes, and right column shows data for $\tau = 3$ minutes. Experiments on different rows were done on different days under the same conditions.  Although there is considerable day-to-day variation of the equilibrium contact angles and damping coefficients, the trends as reported in the main text are consistent for all cases. 
}
\label{extra_water_silane}
\end{figure*}

Figure~\ref{extra_water_silane} shows all three sets of wait-time experiments, all three sets of $\tau \approx 0$ flow-rate experiments, and all three sets of $\tau = 3$ minutes flow-rate experiments for water on silanized glass with best-fit model curves overlaid.  Each row shows data taken on a different days under the same conditions. 
These figures show that although there is considerable day-to-day variation of the equilibrium contact angles and damping coefficients, the trends as to how they vary with wait time, $\tau$, and flow rate, $q$, as reported in the text are consistent for all cases. 

\begin{figure*}
\centering
\includegraphics[width=17.2cm]{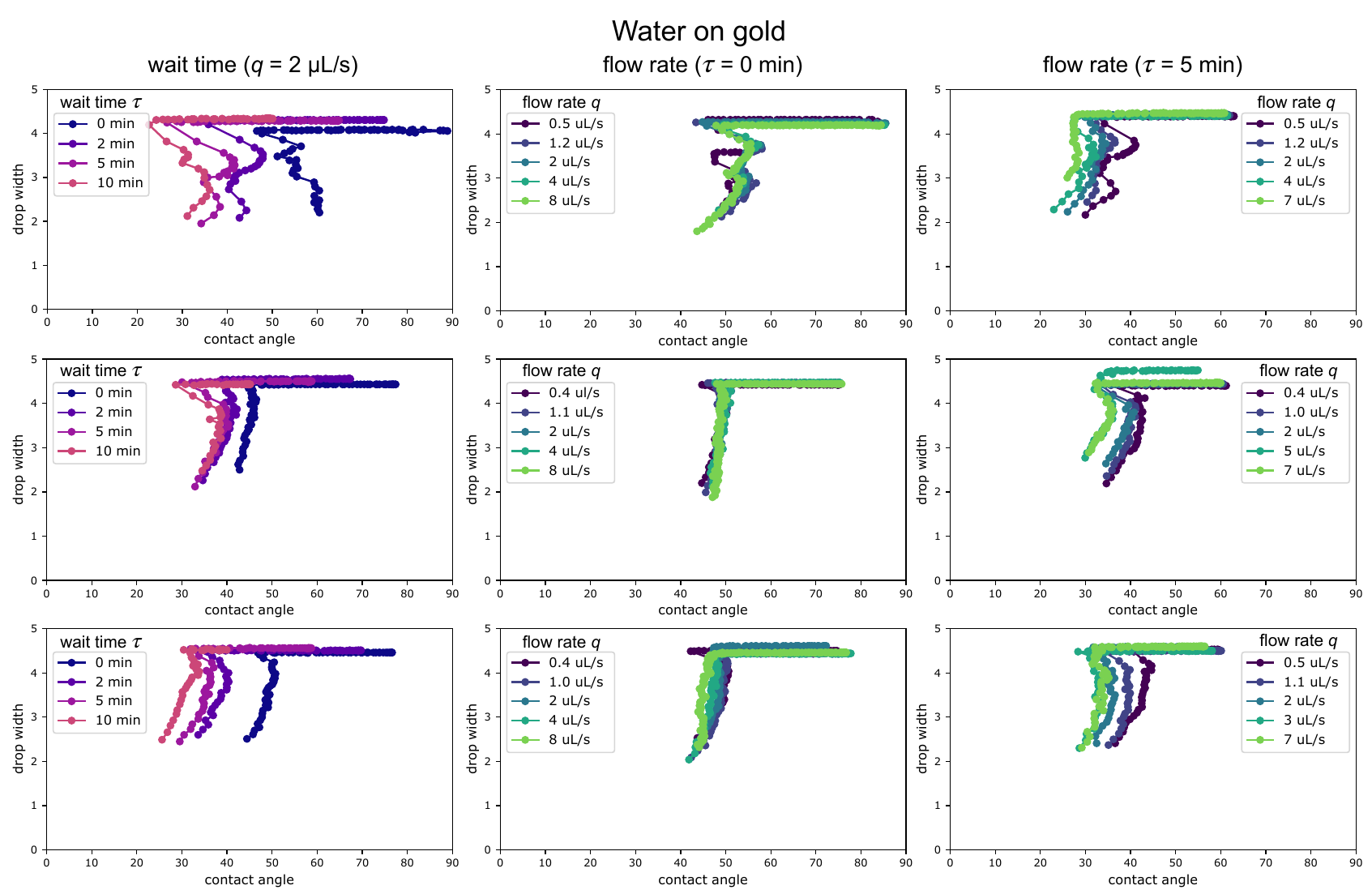}
\caption{
Drop width, $w$ versus contact angle, $\theta$ for water on gold.  Left column shows data at the wait times, $\tau$, in the legend with $q = 2$ $\mu$L/s. The middle and right columns show data at the flow rates, $q$, given in the legend.  Middle column shows data for $\tau \approx 0$ minutes, and right column shows data for $\tau = 5$ minutes. Experiments on different rows were done on different days under the same conditions.  
}
\label{extra_gold}
\end{figure*}

\begin{figure*}
\centering
\includegraphics[width=17.2cm]{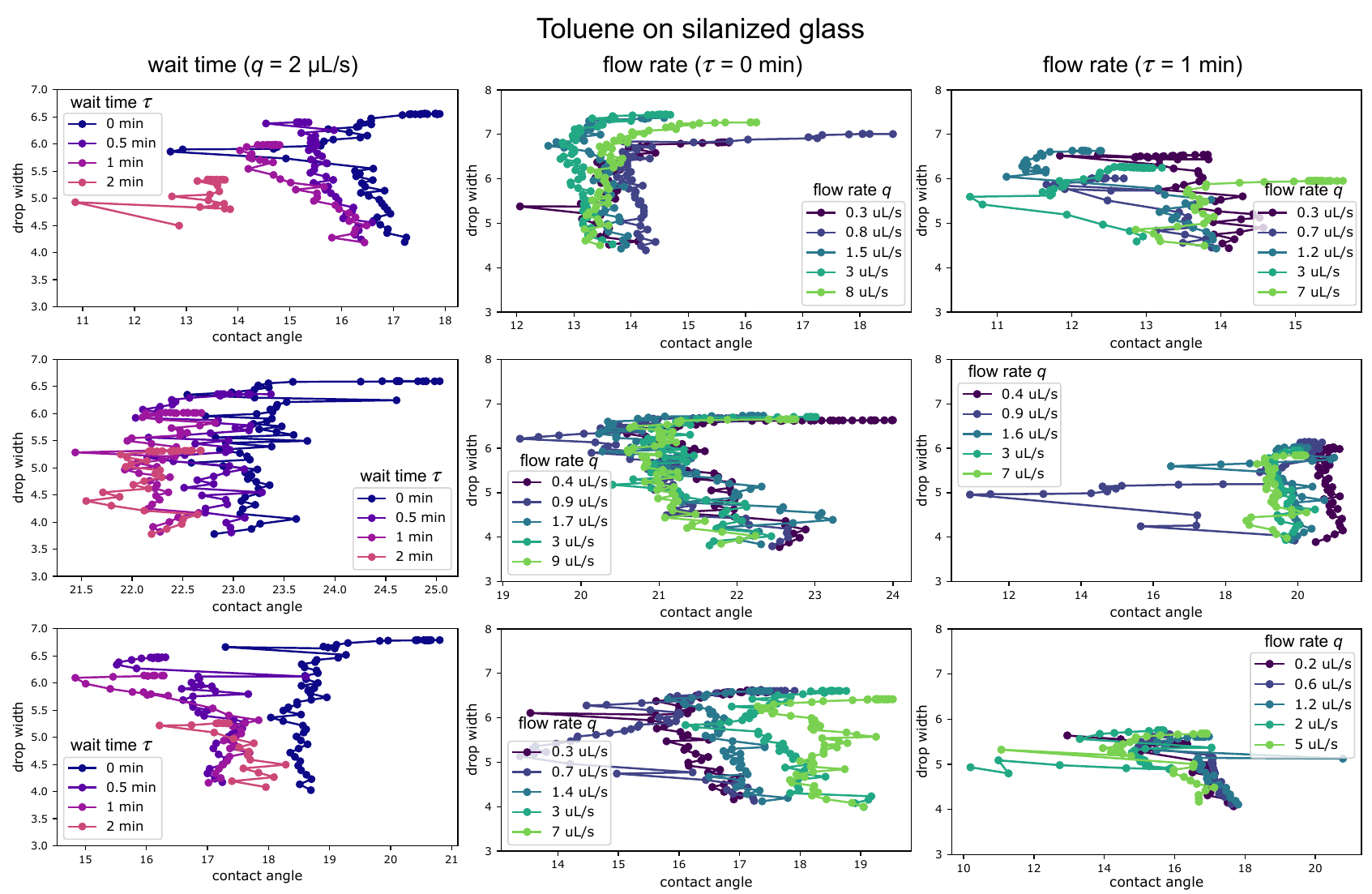}
\caption{
Drop width, $w$ versus contact angle, $\theta$ for toluene on silanized glass.  Left column shows data at the wait times, $\tau$, in the legend with $q = 2$ $\mu$L/s. The middle and right columns show data at the flow rates, $q$, given in the legend.  Middle column shows data for $\tau \approx 0$ minutes, and right column shows data for $\tau = 5$ minutes. Experiments on different rows were done on different days under the same conditions. 
}
\label{extra_toluene}
\end{figure*}

Figure~\ref{extra_gold} shows all experiments as above for water on gold. Fig.~\ref{extra_toluene} shows all experiments for toluene on silanized glass.

\bibliography{main}

\end{document}